\pdfoutput=1
\documentclass[12pt,a4paper]{article}

\usepackage{ifthen} 
\newboolean{pdflatex}
\setboolean{pdflatex}{true} 

\newboolean{articletitles}
\setboolean{articletitles}{true} 

\newboolean{uprightparticles}
\setboolean{uprightparticles}{false} 

\def\paperauthors{LHCb collaboration} 
\def\paperasciititle{Branching fraction measurement of the B0 -> K+ pi- mu+ mu- decay at high dihadron mass} 
\def\papertitle{Branching fraction measurement of the $B^0\to K^+\pi^-\mu^+\mu^-$ decay at high dihadron mass} 
\def\paperkeywords{{High Energy Physics}, {LHCb}} 
\def\papercopyright{\the\year\ CERN for the benefit of the LHCb collaboration} 
\def\paperlicence{CC BY 4.0 licence}
\def\paperlicenceurl{https://creativecommons.org/licenses/by/4.0/}

\newif\ifEnableSectionTOCLinks
\EnableSectionTOCLinksfalse 

\usepackage[top=1in, bottom=1.25in, left=1in, right=1in]{geometry}

\usepackage{microtype}
\usepackage{lineno}  
\usepackage{xspace} 
\usepackage{caption} 

\usepackage{graphicx}  
\usepackage{color}
\usepackage{colortbl}
\graphicspath{{./figs/}} 

\usepackage{amsmath} 
\usepackage{amssymb}
\usepackage{amsfonts}
\usepackage{upgreek} 

\usepackage[normalem]{ulem} 

\newcommand*\patchAmsMathEnvironmentForLineno[1]{%
\expandafter\let\csname old#1\expandafter\endcsname\csname #1\endcsname
\expandafter\let\csname oldend#1\expandafter\endcsname\csname
end#1\endcsname
 \renewenvironment{#1}%
   {\linenomath\csname old#1\endcsname}%
   {\csname oldend#1\endcsname\endlinenomath}%
}
\newcommand*\patchBothAmsMathEnvironmentsForLineno[1]{%
  \patchAmsMathEnvironmentForLineno{#1}%
  \patchAmsMathEnvironmentForLineno{#1*}%
}
\AtBeginDocument{%
\patchBothAmsMathEnvironmentsForLineno{equation}%
\patchBothAmsMathEnvironmentsForLineno{align}%
\patchBothAmsMathEnvironmentsForLineno{flalign}%
\patchBothAmsMathEnvironmentsForLineno{alignat}%
\patchBothAmsMathEnvironmentsForLineno{gather}%
\patchBothAmsMathEnvironmentsForLineno{multline}%
\patchBothAmsMathEnvironmentsForLineno{eqnarray}%
}

\usepackage[pdftex,
            pdfauthor={\paperauthors},
            pdftitle={\paperasciititle},
            pdfkeywords={\paperkeywords}]{hyperref}
\usepackage{hyperxmp}
\hypersetup{
    pdfcopyright={Copyright (C) \papercopyright},
    pdflicenseurl={\paperlicenceurl}
}

\usepackage[colorinlistoftodos,textsize=scriptsize]{todonotes}

\usepackage[bottom,flushmargin,hang,multiple]{footmisc}

\usepackage[all]{hypcap} 

\usepackage{xspace} 
\usepackage{upgreek}

\def\lhcb   {\mbox{LHCb}\xspace}

\def\MagUp {\mbox{\em Mag\kern -0.05em Up}\xspace}

\ifdefined\ifuprightparticles
\else
\newboolean{uprightparticles}
\setboolean{uprightparticles}{false} 
\fi

\ifthenelse{\boolean{uprightparticles}}%
{

 \def\Ppi         {\ensuremath{\uppi}\xspace}

 \def\Pphi        {\ensuremath{\upphi}\xspace}

 \def\Ppsi        {\ensuremath{\uppsi}\xspace}

 \def\PDelta      {\ensuremath{\Delta}\xspace}                 
 \def\PXi         {\ensuremath{\Xi}\xspace}                 
 \def\PLambda     {\ensuremath{\Lambda}\xspace}                 
 \def\PSigma      {\ensuremath{\Sigma}\xspace}                 
 \def\POmega      {\ensuremath{\Omega}\xspace}                 
 \def\PUpsilon    {\ensuremath{\Upsilon}\xspace}
 \let\oldPi\Pi
 \def\PPi         {\ensuremath{\oldPi}\xspace}

 \def\PB      {\ensuremath{\mathrm{B}}\xspace}                 
 \def\PD      {\ensuremath{\mathrm{D}}\xspace}                 
 \def\PJ      {\ensuremath{\mathrm{J}}\xspace}                 
 \def\PK      {\ensuremath{\mathrm{K}}\xspace}                 
 \def\Pb      {\ensuremath{\mathrm{b}}\xspace}

 \def\Pp      {\ensuremath{\mathrm{p}}\xspace}                 

 \def\Ps      {\ensuremath{\mathrm{s}}\xspace}

 \def\thebaroffset{0.0em}
}
{

 \def\Ppi         {\ensuremath{\pi}\xspace}

 \def\Pphi        {\ensuremath{\phi}\xspace}

 \def\Ppsi        {\ensuremath{\psi}\xspace}                 
                  
 \mathchardef\PDelta="7101
 \mathchardef\PXi="7104
 \mathchardef\PLambda="7103
 \mathchardef\PSigma="7106
 \mathchardef\POmega="710A
 \mathchardef\PUpsilon="7107
 \mathchardef\PPi="7105
 \def\PB      {\ensuremath{B}\xspace}                 
 \def\PD      {\ensuremath{D}\xspace}                 
 \def\PJ      {\ensuremath{J}\xspace}                 
 \def\PK      {\ensuremath{K}\xspace}                 
 \def\Pb      {\ensuremath{b}\xspace}

 \def\Pp      {\ensuremath{p}\xspace}                 

 \def\Ps      {\ensuremath{s}\xspace}

 \def\thebaroffset{0.18em}
}
\newcommand{\offsetoverline}[2][\thebaroffset]{\kern #1\overline{\kern -#1 #2}}%

\makeatletter
\ifcase \@ptsize \relax
  \newcommand{\miniscule}{\@setfontsize\miniscule{4}{5}}
\or
  \newcommand{\miniscule}{\@setfontsize\miniscule{5}{6}}
\or
  \newcommand{\miniscule}{\@setfontsize\miniscule{5}{6}}
\fi
\makeatother

\DeclareRobustCommand{\optbar}[1]{\shortstack{{\miniscule (\rule[.5ex]{1.25em}{.18mm})}
  \\ [-.7ex] $#1$}}

\def\squark    {{\ensuremath{\Ps}}\xspace}

\def\bquark    {{\ensuremath{\Pb}}\xspace}

\def\pion   {{\ensuremath{\Ppi}}\xspace}

\def\kaon    {{\ensuremath{\PK}}\xspace}

\def\KorKbar {\kern \thebaroffset\optbar{\kern -\thebaroffset \PK}{}\xspace}

\def\Kstar   {{\ensuremath{\kaon^*}}\xspace}

\newcommand{\phiz}{\ensuremath{\Pphi}\xspace}

\def\D       {{\ensuremath{\PD}}\xspace}

\def\DorDbar {\kern \thebaroffset\optbar{\kern -\thebaroffset \PD}\xspace}

\def\Dp      {{\ensuremath{\D^+}}\xspace}
\def\Dm      {{\ensuremath{\D^-}}\xspace}

\def\DpDm    {\ensuremath{\Dp {\kern -0.16em \Dm}}\xspace}

\def\B       {{\ensuremath{\PB}}\xspace}
\def\Bbar    {{\ensuremath{\offsetoverline{\PB}}}\xspace}

\def\BorBbar {\kern \thebaroffset\optbar{\kern -\thebaroffset \PB}\xspace}
\def\Bz      {{\ensuremath{\B^0}}\xspace}

\def\Bd      {{\ensuremath{\B^0}}\xspace}
\def\Bdb     {{\ensuremath{\Bbar{}^0}}\xspace}
\def\BdorBdbar {\kern \thebaroffset\optbar{\kern -\thebaroffset \Bd}\xspace}
\def\Bu      {{\ensuremath{\B^+}}\xspace}

\def\Bs      {{\ensuremath{\B^0_\squark}}\xspace}

\def\BsorBsbar {\kern \thebaroffset\optbar{\kern -\thebaroffset \Bs}\xspace}

\def\Bds     {{\ensuremath{\B_{(\squark)}^0}}\xspace}

\def\BdorBs  {\Bds}

\def\jpsi     {{\ensuremath{{\PJ\mskip -3mu/\mskip -2mu\Ppsi}}}\xspace}
\def\psitwos  {{\ensuremath{\Ppsi{(2S)}}}\xspace}

\def\Y#1S{\ensuremath{\PUpsilon{(#1S)}}\xspace}

\def\proton      {{\ensuremath{\Pp}}\xspace}
\def\antiproton  {{\ensuremath{\overline \proton}}\xspace}

\def\Lbar        {{\ensuremath{\offsetoverline{\PLambda}}}\xspace}
\def\LorLbar     {\kern \thebaroffset\optbar{\kern -\thebaroffset \PLambda}\xspace}

\def\Lbbar        {{\ensuremath{\Lbar{}^0_\bquark}}\xspace}

\def\BF         {{\ensuremath{\mathcal{B}}}\xspace}

\newcommand{\decay}[2]{\ensuremath{\mathinner{#1\!\to #2}}\xspace}

\def\to                 {\ensuremath{\rightarrow}\xspace}

\def\qsq       {{\ensuremath{q^2}}\xspace}

\def\AT#1     {\ensuremath{A_{\mathrm{T}}^{#1}}\xspace}           

\def\C#1      {\ensuremath{\mathcal{C}_{#1}}\xspace}                       
\def\Cp#1     {\ensuremath{\mathcal{C}_{#1}^{'}}\xspace}                    
\def\Ceff#1   {\ensuremath{\mathcal{C}_{#1}^{\mathrm{(eff)}}}\xspace}        
\def\Cpeff#1  {\ensuremath{\mathcal{C}_{#1}^{'\mathrm{(eff)}}}\xspace}       
\def\Ope#1    {\ensuremath{\mathcal{O}_{#1}}\xspace}                       
\def\Opep#1   {\ensuremath{\mathcal{O}_{#1}^{'}}\xspace}                    

\newcommand{\aunit}[1]{\ensuremath{\text{\,#1}}}       

\newcommand{\tev}{\aunit{Te\kern -0.1em V}\xspace}
\newcommand{\gev}{\aunit{Ge\kern -0.1em V}\xspace}
\newcommand{\mev}{\aunit{Me\kern -0.1em V}\xspace}
\newcommand{\kev}{\aunit{ke\kern -0.1em V}\xspace}
\newcommand{\ev}{\aunit{e\kern -0.1em V}\xspace}
 
\newcommand{\mevc}{\ensuremath{\aunit{Me\kern -0.1em V\!/}c}\xspace}
\newcommand{\gevc}{\ensuremath{\aunit{Ge\kern -0.1em V\!/}c}\xspace}
\newcommand{\mevcc}{\ensuremath{\aunit{Me\kern -0.1em V\!/}c^2}\xspace}
\newcommand{\gevcc}{\ensuremath{\aunit{Ge\kern -0.1em V\!/}c^2}\xspace}
\newcommand{\gevgevcccc}{\ensuremath{\gev^2\!/c^4}\xspace} 

\def\fb   {\ensuremath{\aunit{fb}}\xspace}
\def\invfb   {\ensuremath{\fb^{-1}}\xspace}

\newcommand{\stat}{\aunit{(stat)}\xspace}
\newcommand{\syst}{\aunit{(syst)}\xspace}

\def\deriv {\ensuremath{\mathrm{d}}}

\def\gsim{{~\raise.15em\hbox{$>$}\kern-.85em
          \lower.35em\hbox{$\sim$}~}\xspace}
\def\lsim{{~\raise.15em\hbox{$<$}\kern-.85em
          \lower.35em\hbox{$\sim$}~}\xspace}

\def\sPlot{\mbox{\em sPlot}\xspace}

\def\pt         {\ensuremath{p_{\mathrm{T}}}\xspace}

\def\tell1  {TELL1\xspace}
\def\ukl1   {UKL1\xspace}

\newcommand{\ie}{\mbox{\itshape i.e.}\xspace}

\newcommand{\lhcborcid}[1]{\href{https://orcid.org/#1}{\hspace*{0.1em}\raisebox{-0.45ex}{\includegraphics[width=1em]{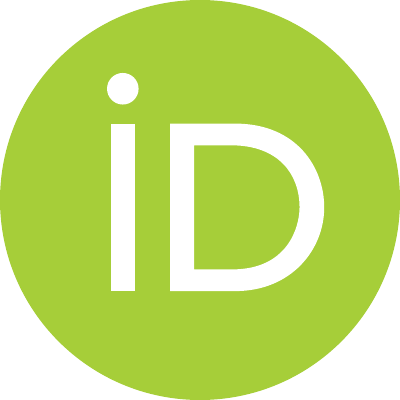}}}}

\hypersetup{
  colorlinks   = true, 
  urlcolor     = blue, 
  linkcolor    = blue, 
  citecolor    = red   
}

\ifEnableSectionTOCLinks
    \usepackage[explicit]{titlesec} 
    
    \let\oldcontentsline\contentsline
    \renewcommand

    \titleformat{\section}{\normalfont\Large\bf}{\hyperlink{tocsection.\thesection}{{\thesection} \parbox[t]{\dimexpr\textwidth-1pc}{#1}}}{1pc}{}

    \titleformat{\subsection}{\normalfont\bf}{\hyperlink{tocsubsection.\thesubsection}{{\thesubsection} \parbox[t]{\dimexpr\textwidth-1pc}{#1}}}{1pc}{}

    \titleformat{name=\section,numberless}[display]{}{}{0pt}{\normalfont\Huge\bfseries #1}
\fi

\usepackage{cite} 
\usepackage{mciteplus}
\makeatletter
\g@addto@macro\bfseries{\boldmath}
\makeatother

\usepackage{longtable} 
\usepackage{subcaption}
\usepackage{booktabs}

\newcommand{\kpimumu}{\decay{\Bz}{K^+\pi^-\mu^+\mu^-}}
\newcommand{\mkpi}{\ensuremath{m_{K\pi}}\xspace}
\newcommand{\mkpimumu}{\ensuremath{m_{K\pi\mu\mu}}\xspace}
\def\Jpsi{\ensuremath{\jpsi}\xspace}
\def\Jpsimm{\decay{\Jpsi}{\mu^+\mu^-}}

\def\btosmumu{\decay{b}{s\mu^+\mu^-}}

\def\btosll{\decay{b}{s\ell^+\ell^-}}

\def\KJpsi{\decay{\Bu}{\Jpsi K^+}}

\def\BdToKpimm{\decay{\Bz}{K^+\pi^-\mu^+\mu^-}}
\def\BdToKpiJpsimm{\decay{\Bz}{\Jpsi(\to\mu^+\mu^-)K^+\pi^-}}
\def\BdToKpiJpsi{\decay{\Bz}{\Jpsi K^+\pi^-}}
\def\BsToKpiJpsi{\decay{\Bs}{\Jpsi K^-\pi^+}}
\def\LbToPKmumu{\decay{\Lbbar}{K^+\antiproton\mu^+\mu^-}}
\def\BsToPiPiMuMu{\decay{\Bs}{\pi^+\pi^-\mu^+\mu^-}}
\def\BsToKKMuMu{\decay{\Bs}{K^+K^-\mu^+\mu^-}}

\def\Kmm{\decay{\Bu}{K^+ \mu^+ \mu^-}}

\def\zfit     {\mbox{\textsc{zfit}}\xspace}
\def\ssyst {\ensuremath{\sigma_{\rm syst}}\xspace}
\def\sstat {\ensuremath{\sigma_{\rm stat}}\xspace}
\def\BsToKPiMuMu{\decay{\Bs}{K^-\pi^+\mu^+\mu^-}}

\begin{document}

\renewcommand{\thefootnote}{\fnsymbol{footnote}}
\setcounter{footnote}{1}


\begin{titlepage}
\pagenumbering{roman}

\vspace*{-1.5cm}
\centerline{\large EUROPEAN ORGANIZATION FOR NUCLEAR RESEARCH (CERN)}
\vspace*{1.5cm}
\noindent
\begin{tabular*}{\linewidth}{lc@{\extracolsep{\fill}}r@{\extracolsep{0pt}}}
\ifthenelse{\boolean{pdflatex}}
{\vspace*{-1.5cm}\mbox{\!\!\!\includegraphics[width=.14\textwidth]{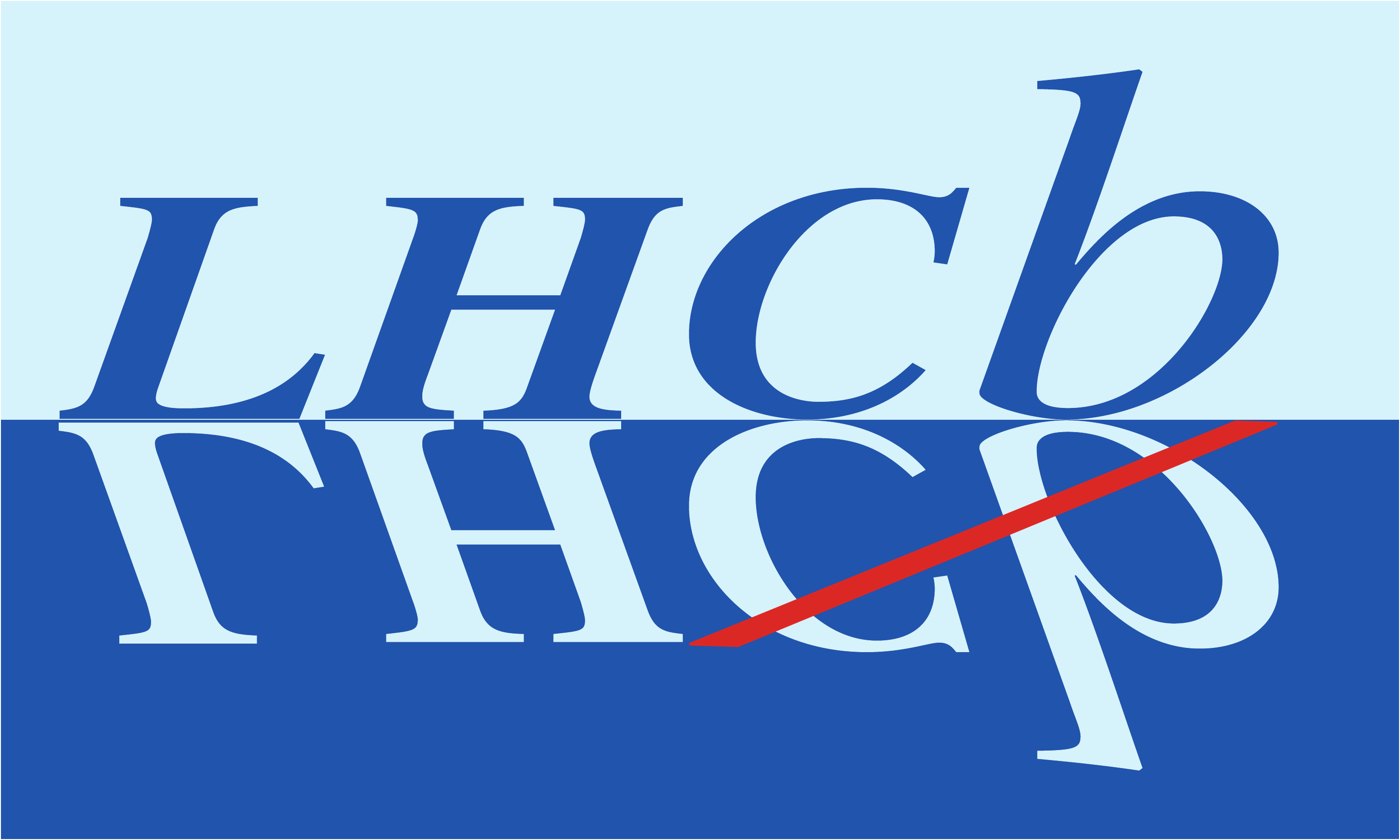}} & &}%
{\vspace*{-1.2cm}\mbox{\!\!\!\includegraphics[width=.12\textwidth]{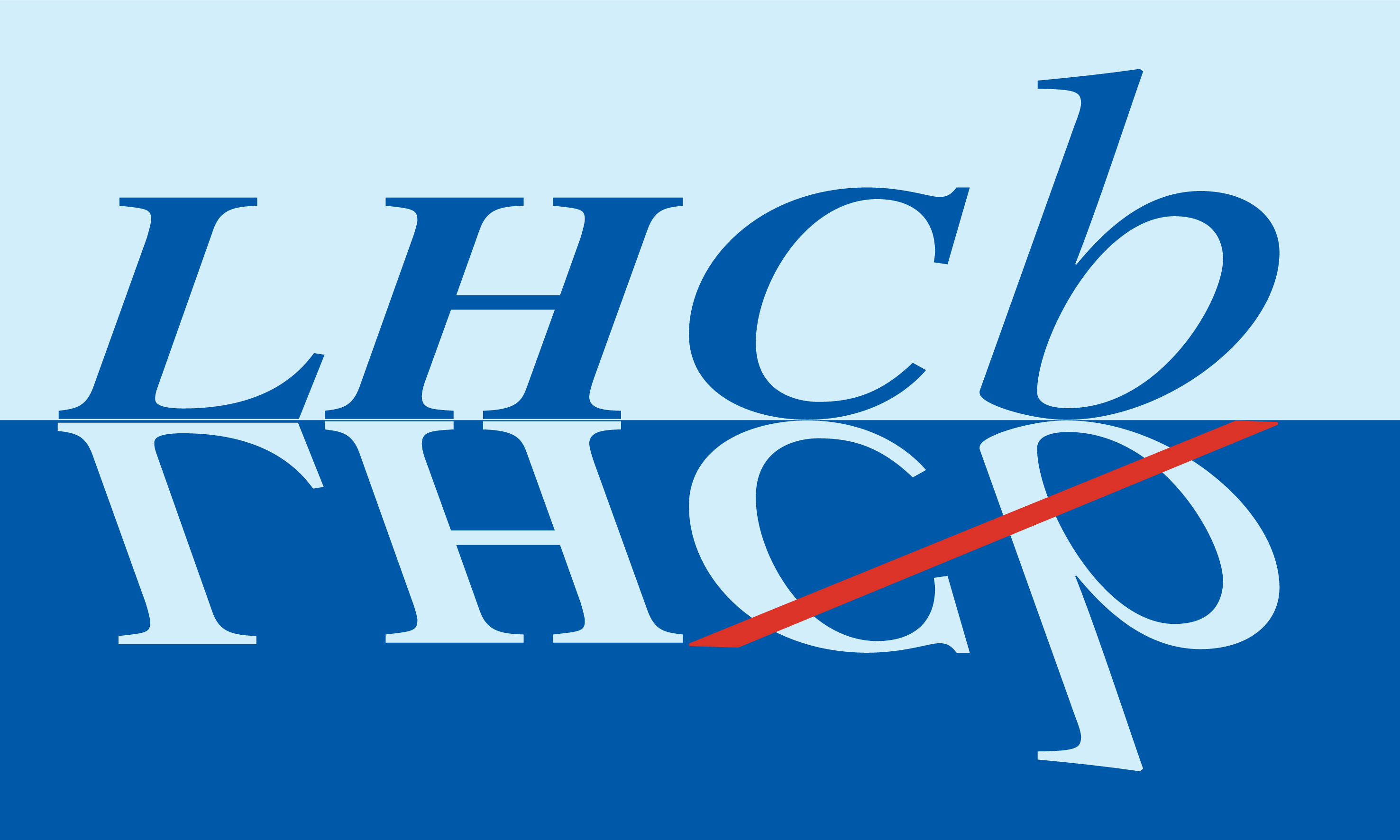}} & &}%
\\
 & & CERN-EP-2026-236 \\  
 & & LHCb-PAPER-2026-005 \\  
 & & October 9, 2026 \\ 
 & & \\
\end{tabular*}

\vspace*{4.0cm}

{\normalfont\bfseries\boldmath\huge
\begin{center}
  \papertitle 
\end{center}
}

\vspace*{1.5cm}

\begin{center}
\paperauthors\footnote{Authors are listed at the end of this Letter.}
\end{center}

\vspace{\fill}

\begin{abstract}
  \noindent
 This Letter presents the first  measurement of the $B^{0} \to K^{+}\pi^{-}\mu^{+}\mu^{-}$ differential branching fraction in the region of the dihadron $K^+\pi^-$ mass, $m_{K\pi}$, 
 \mbox{$1000< m_{K\pi} < 1835\,\text{MeV}/c^{2}$}.
This decay mode is mediated by the quark-level transition
$b \to s\ell^{+}\ell^{-}$, for which the measured effective couplings show tensions with Standard Model predictions. 
The analysis is performed using proton-proton collision data collected with the 
LHCb detector between 2011 and 2018, corresponding to an integrated luminosity of $8.4\,\text{fb}^{-1}$. 
The differential branching fraction is determined across different regions of dimuon 
mass squared, $q^2$, in the interval $0.1 < q^2 < 18.0\,\text{GeV}^2/c^{4}$, as well as in different regions of $ m_{K\pi}$
in the interval $1.1 < q^2 < 6.0\,\text{GeV}^2/c^{4}$.
The results represent the first differential measurement of the branching fraction 
across $m_{K\pi}$ within the selected dihadron-mass region, and the first 
observation of the decay within this region above the $J/\psi$ resonance.

\end{abstract}
\vspace*{2.0cm}

\begin{center}
  Submitted to
  Phys.~Rev.~Lett. 
\end{center}

\vspace{\fill}

{\footnotesize 
\centerline{\copyright~\papercopyright. \href{\paperlicenceurl}{\paperlicence}.}}
\vspace*{2mm}

\end{titlepage}


\newpage
\setcounter{page}{2}
\mbox{~}
%
%
%
%


\renewcommand{\thefootnote}{\arabic{footnote}}
\setcounter{footnote}{0}


\cleardoublepage


\pagestyle{plain} 
\setcounter{page}{1}
\pagenumbering{arabic}


Flavour-changing neutral current processes, such as the quark-level transition \btosll, are strongly suppressed in the Standard Model (SM), making them sensitive to potential contributions from physics beyond the SM.
These contributions would appear as shifts in the Wilson coefficients $C_i$, 
which parameterise the effective couplings in the four-fermion \btosll interaction Hamiltonian. 
Over the past fifteen years, a consistent pattern of deviations has emerged in measurements of both branching fractions and angular observables in several \btosll analyses~\cite{LHCb-PAPER-2015-051,LHCb-PAPER-2016-012,Belle:2016fev,ATLAS:2018gqc,CMS:2024atz,LHCb-PAPER-2025-041,LHCb-PAPER-2014-006,LHCb-PAPER-2020-041,LHCb-PAPER-2021-014,LHCb-PAPER-2021-022,LHCb-PAPER-2024-022,LHCb-PAPER-2025-006}. These deviations can be coherently interpreted as a shift of the vector coefficient $C_9$ from its SM expectation.
One of the most sensitive probes of $C_9$ is the decay \kpimumu,  in which the $K\pi$ system is dominated by the $\Kstar(892)^0$ vector resonance. The inclusion of charge-conjugate processes is implied throughout this Letter. For this reason, existing analyses mainly focus on the region of $K\pi$ invariant masses, \mkpi, below $1050\mevcc$~\cite{LHCb-PAPER-2025-041}. 
The only measurement of the branching fraction beyond this range was performed by the \lhcb experiment using $3\invfb$ of Run~1 data~\cite{LHCb-PAPER-2016-025}. The analysis was restricted to the \mkpi region around the $K^{*}_{0}(1430)^0$ and $K^{*}_{2}(1430)^0$ resonances, $1330 < \mkpi < 1530\mevcc$, and to the \qsq range, $0.1 < q^2 < 8.0\gevgevcccc$, where \qsq denotes the dimuon invariant-mass squared.

Extending the differential branching fraction measurement to the \mkpi region above $1000\mevcc$, and to larger \qsq values,  is important for several reasons.
First,  if the deviations observed in the $\Kstar(892)^0$ region originate from modified short-distance dynamics, similar effects are expected at higher values of $m_{K\pi}$. Furthermore, no observation of \kpimumu decays in this dihadron mass range currently exists for  \qsq regions above the $\jpsi$ resonance.
As discussed in Ref.~\cite{Isidori:2023unk}, semi-inclusive \btosmumu decay rates for $\qsq > 15\gevgevcccc$, obtained by summing exclusive $b\to s\mu^+\mu^-$ measurements,   exhibit deviations similar, though less pronounced, to those observed at lower \qsq values~\cite{LHCb-PAPER-2015-051,LHCb-PAPER-2016-012,Belle:2016fev,ATLAS:2018gqc,CMS:2024atz,LHCb-PAPER-2025-041, LHCb-PAPER-2014-006,LHCb-PAPER-2020-041,LHCb-PAPER-2021-014,LHCb-PAPER-2021-022,LHCb-PAPER-2024-022,LHCb-PAPER-2025-006}.  
Given the phase-space limitations at high $\qsq$,  the $m_{K\pi}$ range  considered in this work, when combined with the results of Ref.~\cite{LHCb-PAPER-2025-041},  represents the first complete determination of the \kpimumu decay width in the $\qsq > 15\gevgevcccc$ region.  This enables a more accurate evaluation of these semi-inclusive rates, in turn helping to provide a complementary probe of the \btosmumu anomalies. Finally, when interpreted in combination with form-factor information from related channels, measurements of the $B^0 \to K^+\pi^-\mu^+\mu^-$ branching fraction over a broader $m_{K\pi}$ range can provide additional constraints on the hadronic inputs entering theory predictions for other exclusive $b\to s\mu^+\mu^-$ processes. In dispersive-based analyses, such information can help reduce the allowed parameter space of the relevant form factors, thereby improving the control of hadronic uncertainties~\cite{Gubernari:2020eft,Gubernari:2023puw,Gubernari:2026sqc}.


In this Letter, the first measurement of the \kpimumu differential branching fraction is presented as a function of both \mkpi and \qsq across the extended kinematic region $1000 < \mkpi < 1835\mevcc$ and $0.1 < \qsq < 18.0\gevgevcccc$.  The measurement is performed differentially in \qsq with interval widths of 1--3\gevgevcccc across the full \qsq and \mkpi range, and differentially in \mkpi with interval widths of 200--300\mevcc within the region $1.1 < \qsq < 6.0\gevgevcccc$.
The measurement is performed using proton-proton ($pp$) collision data recorded by the \lhcb detector during Run~1 (2011--2012) and Run~2 (2016--2018). These data were collected at centre-of-mass energies of 7 and 8\tev for Run~1 and 13\tev for Run~2, corresponding to a total integrated luminosity of 8.4\invfb.

To minimise the dependence on the $b$-hadron production cross-section and reduce systematic uncertainties, the branching fraction measurement is normalised to the \BdToKpiJpsimm decay. 
When measuring differentially across  \mkpi, the branching fraction is extracted  as
\begin{equation}
\begin{split}
    \frac{\deriv^2\BF(\kpimumu)}{\deriv q^2 \deriv\mkpi} = \frac{N_\epsilon(\kpimumu)}{N_\epsilon(\BdToKpiJpsi)} \cdot \frac{\mathcal{B}(\BdToKpiJpsi) \cdot \mathcal{B}(\Jpsimm)}{\Delta q^2\Delta\mkpi} ,
\end{split}
\end{equation}
where the  efficiency-corrected  yields, $N_\epsilon$, 
are obtained from fits to the reconstructed $K^+\pi^-\mu^+\mu^-$ mass distribution, \mkpimumu, after weighting each candidate for efficiency effects as determined from simulation.
The quantities $\Delta q^2$ and $\Delta\mkpi$ correspond to the width of \qsq and \mkpi intervals considered.  When performing the measurement integrated over the range $1000 < \mkpi < 1835\mevcc$, the division by $\Delta\mkpi$ is omitted. The \BdToKpiJpsi yield is measured in the extended region $796< \mkpi< 1835\mevcc$, as the corresponding branching fraction measurement, taken from Ref.~\cite{Belle:2014nuw},  has a  smaller uncertainty than the equivalent value measured in the range of $1000 < \mkpi < 1835\mevcc$ used for the \kpimumu mode.  

The LHCb detector is a single-arm forward spectrometer covering the pseudorapidity range $2 < \eta < 5$, 
described in detail in Refs.~\cite{LHCb-DP-2008-001,LHCb-DP-2014-002}. 
The online event selection uses a hardware trigger based on the muon \pt,
with a threshold ranging from 1.5 to 2.9\gevc depending on the data-taking period.
This is followed by a software trigger~\cite{LHCb-DP-2012-004},
which requires a high-\pt track significantly displaced from every primary
vertex (PV). This track must form a good-quality vertex with at least one
other track, with candidates satisfying additional topological
criteria~\cite{BBDT,LHCb-PROC-2015-018}.


Large samples of simulated events, produced using the software described in 
Refs.~\cite{Sjostrand:2007gs,Lange:2001uf,Allison:2006ve,*Agostinelli:2002hh}, 
are used to determine reconstruction and selection efficiencies and to estimate residual background contamination.
Inaccuracies in the simulation are corrected using control samples from data. 
The simulated particle identification~(PID) response is corrected using dedicated  control samples~\cite{LHCb-PUB-2016-021,LHCb-DP-2018-001}, and the track-reconstruction efficiency is corrected using methods described in Ref.~\cite{LHCb-DP-2013-002}. 
The mismodelling in simulation of the \Bz momentum, transverse momentum~($p_\mathrm{T}$), vertex-fit quality, and impact-parameter $\chi^2$, which quantifies the change in the PV fit quality when the track of interest is included and excluded, is corrected using background-subtracted \BdToKpiJpsi decays weighted with the \texttt{GBReweighter} algorithm~\cite{Rogozhnikov:2016bdp}.
The event multiplicity is corrected independently using a one-dimensional histogram-based weighting 
of the number of reconstructed tracks.
Corrections to the hardware-trigger response are derived from data using \KJpsi decays~\cite{LHCb-PUB-2014-039} as a function of the largest $p_\mathrm{T}$ of the two muons.

Signal candidates are reconstructed from a pair of oppositely charged muons combined with a $K^+\pi^-$ pair. All final-state tracks are required to have a significant impact parameter with respect to any PV 
and to form a common vertex of good fit quality. 
The \Bz candidate must be consistent with originating from one of the PVs, 
whilst having a decay vertex which is significantly displaced from it. 
The angle between the reconstructed \Bz momentum and the line connecting the PV and decay vertex is required to be small. 
All final-state tracks are required to have PID information consistent with their mass assignments.

Several sources of background remain  after applying these selection criteria.  These include other \btosmumu decays, which can contribute to the background if one or more hadrons are misidentified. 
To reduce these to negligible levels, additional PID vetoes are applied when the invariant mass of a candidate, 
under specific particle-mass hypotheses, lies near the known mass of a potential background decay. 
Specific vetoes are applied against \LbToPKmumu, \BsToPiPiMuMu and \BsToKKMuMu decays,  where a  single hadron is misidentified,  as well as against $\Bdb\to K^-\pi^+\mu^+\mu^-$ decays, 
where the kaon and pion are misidentified as each other. 
Charmonium decays of the type $\decay{\BdorBs}{\Jpsi h^+h^-}$, where $h^{\pm}$ is a charged hadron, can also constitute a background if one of the muons is misidentified as a hadron, and vice versa. The dominant contribution arises from \BdToKpiJpsi decays, 
where $K\pi$ combinations across the full kinematically allowed $m_{K\pi}$ range can contribute. 
These are reduced to negligible levels by 
vetoing candidates that, when assigning the hadron candidate a muon mass,  result in a \qsq value close to that of the $\jpsi$ or $\psitwos$ resonances.  
To suppress contributions from leading order, \ie tree-level,  transitions, regions of the \qsq spectrum near the $\phiz(1020)$, $\Jpsi$, and $\psitwos$ resonances are excluded.
Candidates satisfying $9.0<\qsq<10.2 \gevgevcccc$ and $796<\mkpi<1835 \mevcc$ are selected for the normalisation channel. 

To suppress background from coincidental track combinations, 
known as combinatorial background, a boosted decision tree (BDT) classifier is trained. 
The BDT uses simulated \BdToKpimm decays as a proxy for signal, while data candidates with 
$5500 <\mkpimumu <7000  \mevcc$ are used as a proxy for background. 
The BDT is trained using variables describing the $B^0$ candidate's momentum,  flight distance and compatibility with the associated PV, as well as the decay-vertex quality and PID information of the final-state particles. The requirement on the BDT response is optimised to maximise the expected signal significance.  To avoid experimenter's bias, the results of the analysis were not examined until the full selection procedure had been finalised.

The geometry of the detector and the applied selection remove signal candidates unevenly across the phase space of the decay. 
Since the signal decay amplitude in the high-$m_{K\pi}$ region receives contributions from multiple overlapping partial waves,  the efficiency is parameterised as a function of all five kinematic variables 
needed to describe the decay, namely  \qsq, \mkpi and three decay angles.  The three decay angles used are the lepton helicity angle, $\cos\theta_\ell$, the hadron helicity angle, $\cos\theta_K$, and the angle between the planes formed by the hadron and the lepton decay products, $\phi$. These angles are collectively denoted by $\Omega$, and their complete definition can be found in Ref.\cite{LHCb-PAPER-2013-019}.

This five-dimensional efficiency is parameterised as a sum of products of Legendre polynomials and is applied to both the \kpimumu and \BdToKpiJpsi mode.
A good efficiency  description is achieved by considering separate functions defined across three distinct   \qsq regions, namely 
$0.08 < q^2 < 1.10\gevgevcccc$ (low-\qsq),  $1.1 < q^2 < 12.5\gevgevcccc$ (central-\qsq), and $12.5 < q^2 < 18.3\gevgevcccc$ (high-\qsq). 
The low-\qsq and high-\qsq efficiency functions extend beyond the \qsq analysis region to avoid edge effects. These functions account only for relative efficiency variations, as the absolute efficiency scale cancels between the \kpimumu and \BdToKpiJpsi mode. The efficiency for the requirement that candidates be in a given \mkpimumu range, and that \BdToKpiJpsi candidates be in a given \qsq range, are applied separately. 
The efficiency function is determined using a combination of Run~1 and Run~2 simulation,  where the relative contribution between Run~1 and Run~2 is fixed according to the equivalent  ratio of  observed efficiency-corrected yields,  $N_\epsilon(\BdToKpiJpsi)$,  denoted $r_{Y}$.  
The $\phi$ angle is not included in the low-\qsq efficiency parameterisation, as no significant $\phi$ dependence is observed. Including all five dimensions is also found to reduce the stability of the parameterisation due to the limited simulation sample size available at low \qsq.

\begin{figure}[tb]
    \centering
     
    \includegraphics[width=0.49\textwidth]{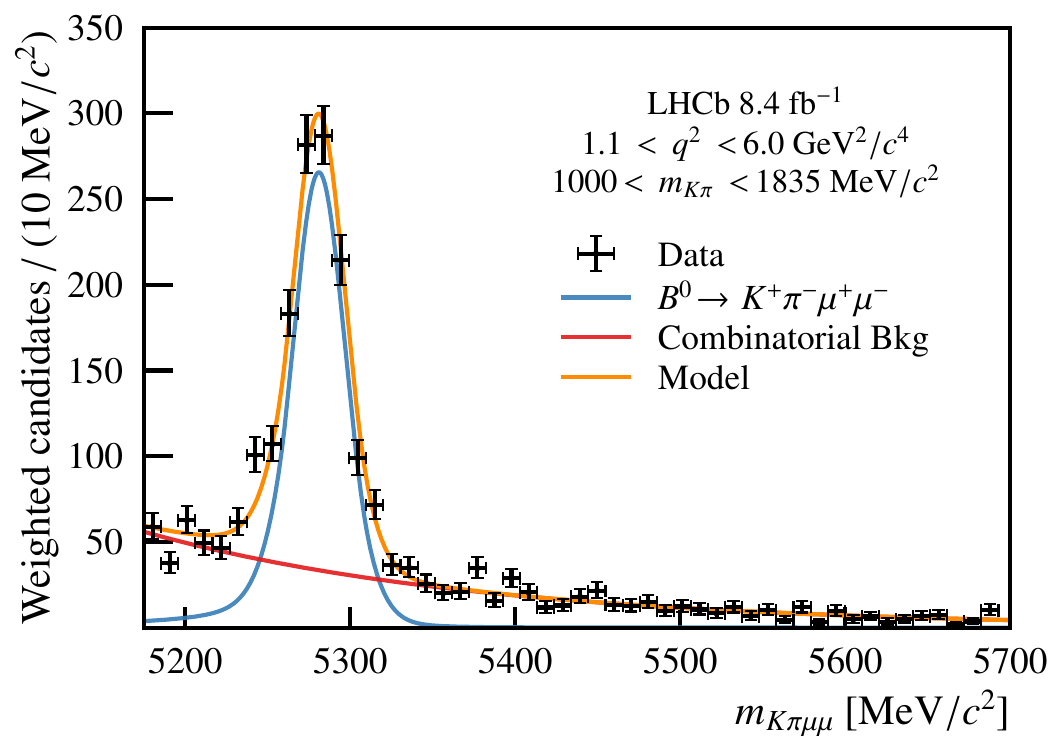}
    \includegraphics[width=0.49\textwidth]{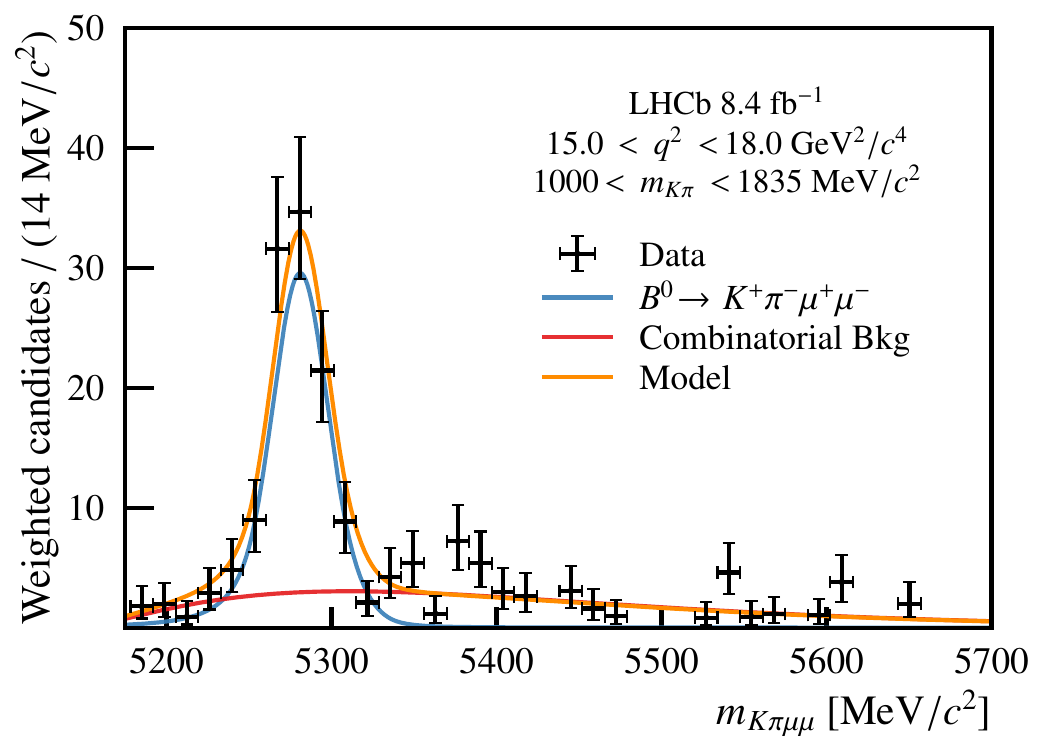}
    \caption{Efficiency-corrected mass distributions and fits to the $\decay{B^0}{K^+\pi^-\mu^+\mu^-}$ candidates in two different regions of the \qsq spectrum. }
    \label{fig:FitProjections-1}
\end{figure}

The signal and control-channel yields are extracted via unbinned weighted maximum-likelihood fits to the four-body invariant mass in the region $5175 < \mkpimumu< 5700 \mevcc$  using the \zfit package~\cite{Eschle:2019jmu}, with weights given by the inverse of the five-dimensional efficiency. 
The signal lineshape in \mkpimumu is described  using the sum of Run~1 and Run~2 models, combined according to the yield ratio $r_Y$. Each model comprises the sum of a Gaussian function and a double-sided Crystal Ball (DCB) 
function~\cite{Skwarnicki:1986xj} with a shared peak position, where the DCB tail parameters are fixed to the values observed in simulation. For the normalisation mode, the peak position and widths of the Gaussian and DCB functions are determined from fits to data, while for the signal they are fixed to the values observed in the normalisation mode. An additional width scaling factor is applied for the signal mode to account for the different resolution observed between the signal and normalisation mode in the $1000 < m_{K\pi} < 1835\mevcc$ window, and across different \qsq regions. 
The only background component considered in the fit to the signal candidates is the combinatorial background, parameterised with a decreasing exponential function with a freely floating slope parameter.  
This description needs to be modified in the high-\qsq region to account for distortion induced by the reduced available phase space, and follows a similar approach to that described in Ref.~\cite{LHCb-PAPER-2024-056}. Example fits to the \kpimumu mass spectra are shown in Fig.~\Ref{fig:FitProjections-1}. The fit to the normalisation mode includes an additional background component for the decay \BsToKpiJpsi, which shares the parameterisation of the \Bz mode, but with the peak position shifted to take into account the mass difference between \Bz and \Bs mesons~\cite{PDG2026}. The possible effects of the analogous \Bs channel on the signal fit are considered as a source of systematic uncertainty.

\begin{figure}[tb]
    \centering
    \includegraphics[width=0.49\textwidth]{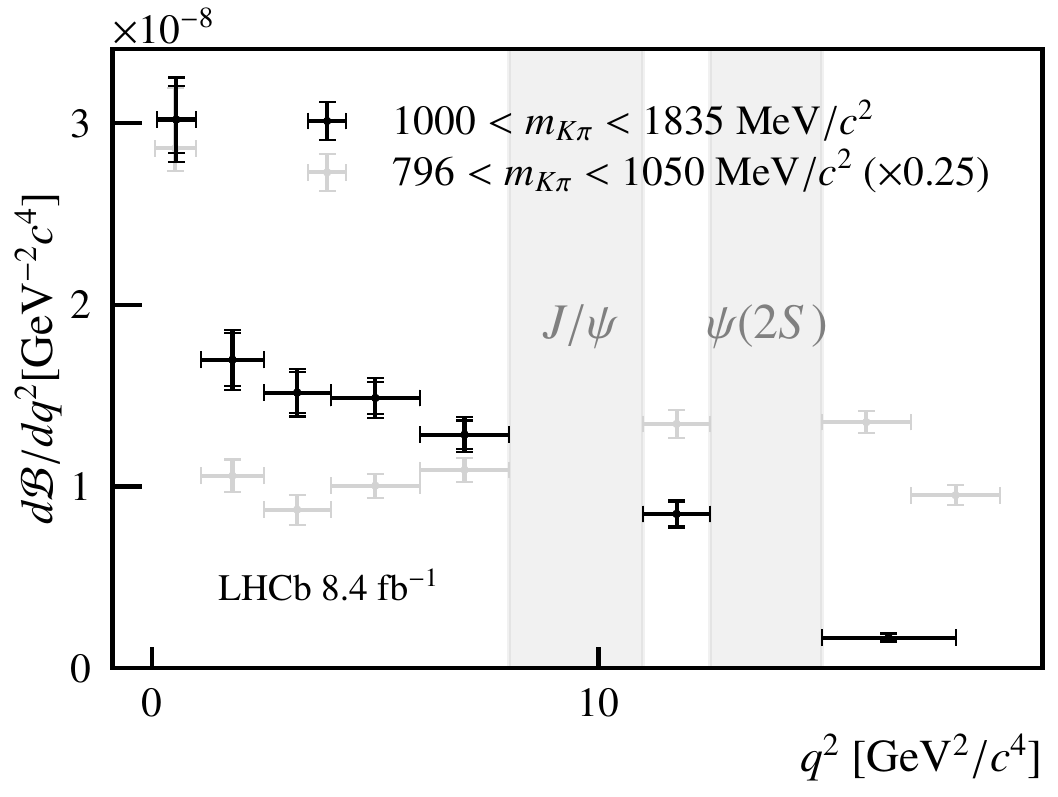}
    \includegraphics[width=0.49\textwidth]{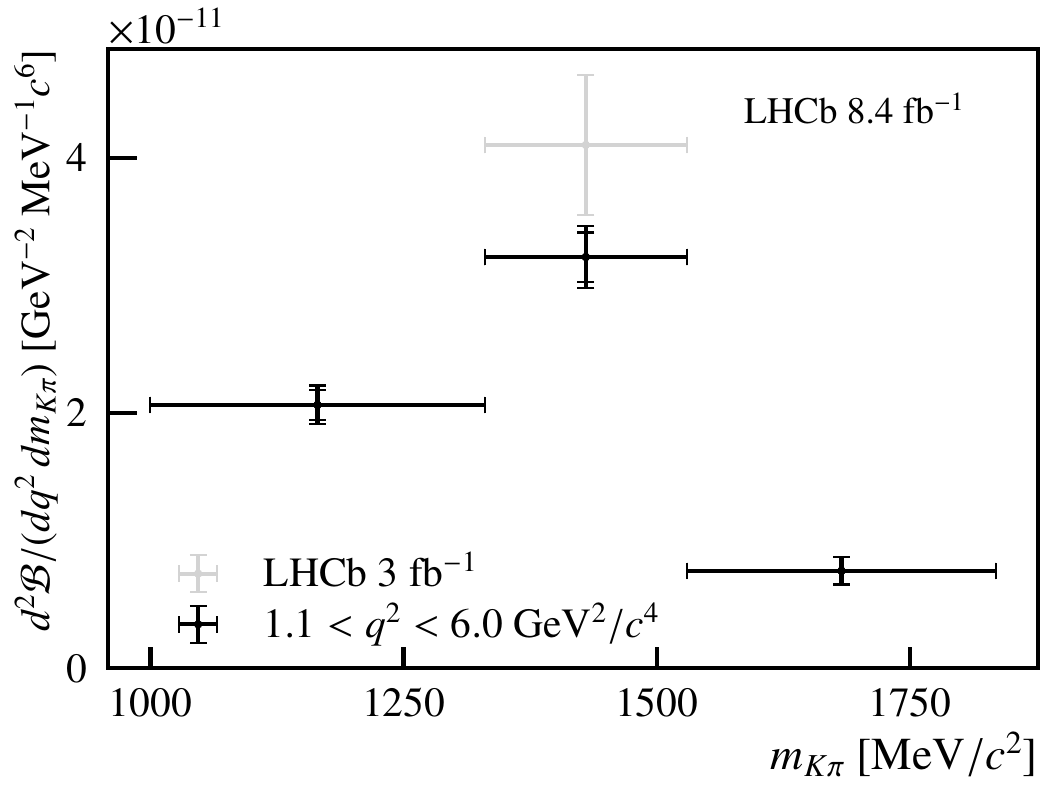}
    \caption{
    Differential branching fraction of the decay
     $B^0 \to K^+\pi^-\mu^+\mu^-$  as a function of (left)~\qsq and (right)
    $m_{K\pi}$ in the central-\qsq region.  
    On the left, the light grey data points
    show the differential branching fraction of the 
    $B^0 \to K^+\pi^-\mu^+\mu^-$ decay measured in the region
    $796<m_{K\pi}<1050\mevcc$,  obtained from 
    Ref.~\cite{LHCb-PAPER-2025-041} and scaled by 0.25 to allow easier comparison with the result of this Letter.  On the right, the light grey data point
    corresponds to the branching fraction taken from
    Ref.~\cite{LHCb-PAPER-2016-025}, also  integrated over
    $1.1<q^2<6.0\gevgevcccc$.  All other data points show the branching fraction extracted in this work and have two error bars, where the inner error bar represents  the statistical and systematic
    uncertainty excluding  the uncertainty on the branching
    fraction of the normalisation mode.
    }
    \label{fig:result}
\end{figure}

Several sources of systematic uncertainties~(\ssyst) are evaluated, the largest of which comes from limited knowledge of the branching fraction of the normalisation mode \BdToKpiJpsi in the region $796 < m_{K\pi} < 1835\mevcc$, which is calculated using the measurement performed in Ref.~\cite{Belle:2014nuw}. Pseudoexperiments based on the amplitude model provided in Ref.~\cite{Belle:2014nuw} are used to deduce the fraction of \BdToKpiJpsi candidates in this \mkpi region, which is found to be $0.966\pm0.013$. This yields 
\begin{equation}
    \mathcal{B}(B^0\to \jpsi(\to\mu^+\mu^-) \kaon^+\pion^- ,796 < m_{K\pi} < 1835\mevcc) = 6.6 \pm 0.3\times 10^{-5} 
\end{equation}

The total 4.6\% systematic uncertainty due to the limited knowledge of $\mathcal{B}(\BdToKpiJpsi)$ corresponds to between 39\% and 123\% of the statistical uncertainty~(\sstat) depending on the  measurement region in \qsq and \mkpi. 
 
The next largest systematic uncertainty comes from the order of polynomial used to describe the selection efficiency. To account for this, the polynomial order is increased by one in each of the five dimensions. The data is bootstrapped with replacement~\cite{efron:1979} and refitted twice, once with weights from the baseline efficiency function and once with weights from the alternative function. The systematic uncertainty is taken as the quadratic sum of the mean and width of the distribution of differences from the bootstrapped samples.   The result ranges from 11\% to 44\% of \sstat, depending on the region. Systematic variations that probe changes in the efficiency-function parameterisation are estimated with data in order to avoid any dependence on an assumed decay model. 

Other sizeable systematic uncertainties include effects due to: the limited amount of simulation used to parameterise the efficiency functions, estimated by resampling the simulation samples and using these alternative efficiencies to perform a fit to data; the presence of any residual background contamination, estimated from pseudoexperiments generated with realistic contamination and fitted with and without accounting for those additional contributions;  the presence of any \BsToKPiMuMu contamination; and an alternative scheme~\cite{LHCb-PAPER-2022-046,LHCb-PAPER-2022-045} to correct for differences between simulation and data. 
Sources of systematic uncertainty arising from the choice of signal fit model contribute less than 10\% of \sstat. These are quantified by: using an alternative parameterisation of the mass peak,  given by the sum of two DCB functions; applying a looser definition of what constitutes a signal event in simulation; and considering different yield ratios, $r_Y$, when averaging the Run~1 and Run~2 distributions. A full summary of the systematic uncertainty sources is given in Table~\ref{tab:syst}.



\begin{table}[tb]
\centering
\caption{Relative systematic uncertainties on the differential branching fraction. Ranges indicate the variation across the \qsq\ and \mkpi\ measurement intervals. The uncertainty from the branching fraction of the normalisation mode, ${\mathcal B}(\decay{\Bd}{\jpsi K^+\pi^-})$, is quoted separately.
}
\label{tab:systs_suppl}
\begin{tabular}{lcc}\toprule

Source & $\ssyst/\frac{\deriv\mathcal{B}}{\deriv \qsq}$ & $\ssyst/\frac{\deriv^2\mathcal{B}}{\deriv \qsq\deriv \mkpi}$\\ 
\bottomrule
Higher-order efficiency          & $0.007\text{--}0.031$  & $0.008\text{--}0.052$ \\ 
Simulation sample size           & $0.004\text{--}0.024$ & $0.004\text{--}0.029$ \\ 
Residual backgrounds             & $0.003\text{--}0.029$ & $0.004\text{--}0.021$ \\ 
\Bs background                 & $0.005\text{--}0.015$ & $0.011\text{--}0.024$ \\ 
Alternative correction scheme    & $0.001\text{--}0.016$ & $0.005\text{--}0.011$ \\ 
Signal fit model                 & $0.002\text{--}0.010$  & $0.003\text{--}0.005$ \\ 
\midrule
Quadratic sum                    & $0.011\text{--}0.047$   & $0.017\text{--}0.069$  \\
 \midrule

Normalisation ${\mathcal B}(\decay{\Bd}{\jpsi K^+\pi^-})$
& \multicolumn{1}{c}{$0.046$}
& \multicolumn{1}{c}{$0.046$} \\
\bottomrule
\end{tabular}
\label{tab:syst}
\end{table}

 Multiple cross-checks are performed to assess the robustness of the result against: tightening the PID requirement 
 on all final-state particles; tightening the combinatorial BDT requirement such that the background is reduced by $\sim$50\% with respect to the baseline requirement;  extracting the branching fraction using the \sPlot technique~\cite{Pivk:2004ty} to subtract background instead of fitting it directly; reducing the fit range in \mkpimumu; including additional corrections to simulation on the hadron $p$ and $p_\mathrm{T}$; and parameterising the efficiency of the \mkpimumu selection directly in the acceptance function.  All checks result in shifts with an absolute size less than \sstat. 
For all cross-checks which deviate by more than 0.5\sstat it is checked that, when taking correlations into account, they are compatible with the original measurement at the level of $2.5$ standard deviations or better.  
 
 
The results of the branching fraction measurements are shown in Fig.~\ref{fig:result} and reported in Table~\ref{tab:BF_results_q2} and~\ref{tab:BF_results_mkpi} in the End Matter. In Fig.~\ref{fig:result}\,(left), the \kpimumu branching fraction in the region $796<\mkpi<1050$\mevcc, calculated using the measurements performed on 8.4\invfb of data provided in Ref.~\cite{LHCb-PAPER-2025-041}, is shown. In Fig.~\ref{fig:result}\,(right) a result from Ref.~\cite{LHCb-PAPER-2016-025}  is shown.  
The signal significance is determined from the  $p$-value of the background-only hypothesis, obtained using  Wilks’ theorem. The significance is well above five standard deviations in every region. 
There are currently no SM predictions for the \kpimumu decay in  the dihadron  mass range considered in this paper. The notable decrease in the branching fraction above the \psitwos region in \qsq can be attributed to the decreasing decay phase space. 

In summary, this Letter presents the first measurement of the differential branching fraction of the \kpimumu decay as a function of \qsq in the region \mbox{$0.1 < \qsq < 18.0\gevgevcccc$} and $1000 < \mkpi < 1835\mevcc$, and as a function of \mkpi in the region \mbox{$1.1 < \qsq < 6.0\gevgevcccc$} and \mbox{$1000 < \mkpi < 1835\mevcc$}. 
This measurement is compatible with, and supersedes,  the branching fraction determination previously obtained by the \lhcb collaboration using Run~1 data~\cite{LHCb-PAPER-2016-025}. 
The analysis further provides the first observation of \kpimumu decays in the dihadron mass range $1000 < \mkpi < 1835\mevcc$ in two regions of $\qsq$ above the $\jpsi$ resonance.   Theory predictions for $\mathcal{B}(\kpimumu)$ in the \mkpi region above $1000\mevcc$ are needed in order to assess the compatibility of these results  with the SM. 
Measurements across the full \qsq region are particularly valuable, as the size of nonlocal hadronic effects may differ between low- and high-$\qsq$ regions~\cite{Isidori:2025dkp}.
Combined with the measurement discussed in Ref.~\cite{LHCb-PAPER-2025-041}, this work completes the determination of the \kpimumu decay width.  It also enables an updated evaluation of the semi-inclusive experimental \btosmumu rate when considered together with the results on \Kmm decays~\cite{LHCb-PAPER-2014-006},  thereby improving the comparison with theoretical predictions, particularly in the high-$\qsq$ region~\cite{Isidori:2023unk}. 

\section*{Acknowledgements}
%
%
\noindent We express our gratitude to our colleagues in the CERN
accelerator departments for the excellent performance of the LHC. We
thank the technical and administrative staff at the LHCb
institutes.
We acknowledge support from CERN and from the national agencies:
ARC (Australia);
CAPES, CNPq, FAPERJ and FINEP (Brazil); 
MOST and NSFC (China); 
CNRS/IN2P3 and CEA (France);  
BMFTR, DFG and MPG (Germany);
NKFIH (Hungary);              
INFN (Italy); 
NWO (Netherlands); 
MNiSW and NCN (Poland); 
MEC/IFA (Romania); 
MICIU and AEI (Spain);
SNSF and SER (Switzerland); 
NASU (Ukraine); 
STFC (United Kingdom); 
DOE NP and NSF (USA).
We acknowledge the computing resources that are provided by ARDC (Australia), 
CBPF (Brazil),
CERN, 
IHEP and LZU (China),
IN2P3 (France), 
KIT and DESY (Germany), 
INFN (Italy), 
SURF (Netherlands),
Polish WLCG (Poland),
IFIN-HH (Romania), 
PIC (Spain), CSCS (Switzerland), 
GridPP (United Kingdom),
and NSF (USA).  
We are indebted to the communities behind the multiple open-source
software packages on which we depend.
Individual groups or members have received support from
RTP (Australia), 
FWO Odysseus grant G0ASD25N (Belgium), 
Key Research Program of Frontier Sciences of CAS, CAS PIFI, CAS CCEPP (China); 
Minciencias (Colombia);
EPLANET, Marie Sk\l{}odowska-Curie Actions, ERC and NextGenerationEU (European Union);
A*MIDEX, ANR, IPhU and Labex P2IO, and R\'{e}gion Auvergne-Rh\^{o}ne-Alpes (France);
Alexander-von-Humboldt Foundation (Germany);
ICSC (Italy); 
Severo Ochoa and Mar\'ia de Maeztu Units of Excellence, GVA, XuntaGal, GENCAT, InTalent-Inditex and Prog.~Atracci\'on Talento CM (Spain);
the Leverhulme Trust, the Royal Society and UKRI (United Kingdom).

\clearpage 

\clearpage
\section*{End matter}
\renewcommand{\thesubsection}{\Alph{subsection}}
\subsection{Branching fraction results}
\label{AppBFresults}
Tables~\ref{tab:BF_results_q2} and \ref{tab:BF_results_mkpi} show the values of the differential branching fractions measured as a function of $q^2$ and \mkpi, respectively. 
\begin{table}[!h]
    \centering
    \caption{Differential branching fraction \mbox{$\deriv {\mathcal B}(\decay{\Bd}{K^+\pi^-\mu^+\mu^-})/\deriv q^2$} in intervals of $q^2$. The uncertainties are, in order, statistical, systematic, and due to the uncertainty on the branching fraction of the normalisation mode.}
    \begin{tabular}{r c}
    \hline
    $q^2$ [\gevgevcccc] & $\deriv\mathcal{B}/\deriv q^2 \times 10^{-8}[\mathrm{GeV}^{-2}c^4]$ \\ \hline \hline
    0.10 -- 0.98 & $3.02 \pm 0.18\stat \pm 0.03\syst \pm 0.13$\,(ext) \\ 
    1.1 -- 2.5 & $1.70 \pm 0.12\stat \pm 0.07\syst \pm 0.07$\,(ext) \\ 
    2.5 -- 4.0 & $1.52 \pm 0.11\stat \pm 0.03\syst \pm 0.07$\,(ext) \\ 
    4.0 -- 6.0 & $1.49 \pm 0.08\stat \pm 0.02\syst \pm 0.07$\,(ext) \\ 
    6.0 -- 8.0 & $1.29 \pm 0.08\stat \pm 0.02\syst \pm 0.06$\,(ext) \\ 
    11.0 -- 12.5 & $0.85 \pm 0.06\stat \pm 0.02\syst \pm 0.04$\,(ext) \\ 
    15.0 -- 18.0 & $0.17 \pm 0.02\stat \pm 0.01\syst \pm 0.01$\,(ext) \\ \hline
    1.1 -- 6.0 & $1.56 \pm 0.06\stat \pm 0.03\syst \pm 0.07$\,(ext) \\ \hline
    \end{tabular}
    \label{tab:BF_results_q2}
\end{table}

\begin{table}[!h]
    \centering
    \caption{Differential branching fraction \mbox{$\deriv {\mathcal B}(\decay{\Bd}{K^+\pi^-\mu^+\mu^-})/\deriv q^2\deriv m_{K\pi}$} in intervals of $m_{K\pi}$ in the central-$q^2$ region, $1.1<q^2<6.0\gevgevcccc$. The uncertainties are, in order, statistical, systematic, and due to the uncertainty on the branching fraction of the normalisation mode.}
    \begin{tabular}{c c}
    \hline
    $m_{K\pi}$ [\mevcc]  & $\deriv^2 \mathcal{B}/(\deriv q^2\deriv m_{K\pi}) \times 10^{-11}[\mathrm{GeV}^{-2}\mathrm{MeV}^{-1}c^6]$\\ \hline \hline
    1000 -- 1330 & $2.06 \pm 0.11\stat \pm 0.03\syst \pm 0.09$\,(ext) \\
    1330 -- 1530 & $3.22 \pm 0.18\stat \pm 0.06\syst \pm 0.14$\,(ext) \\ 
    1530 -- 1835 & $0.76 \pm 0.09\stat \pm 0.05\syst \pm 0.03$\,(ext) \\ \hline
    1000 -- 1835 & $1.87 \pm 0.07\stat \pm 0.03\syst \pm 0.08$\,(ext) \\ \hline
    \end{tabular}
    \label{tab:BF_results_mkpi}
\end{table}


\subsection{Examples of data distributions}
\begin{figure}[tb]
    \centering
    \includegraphics[width=0.52\textwidth]{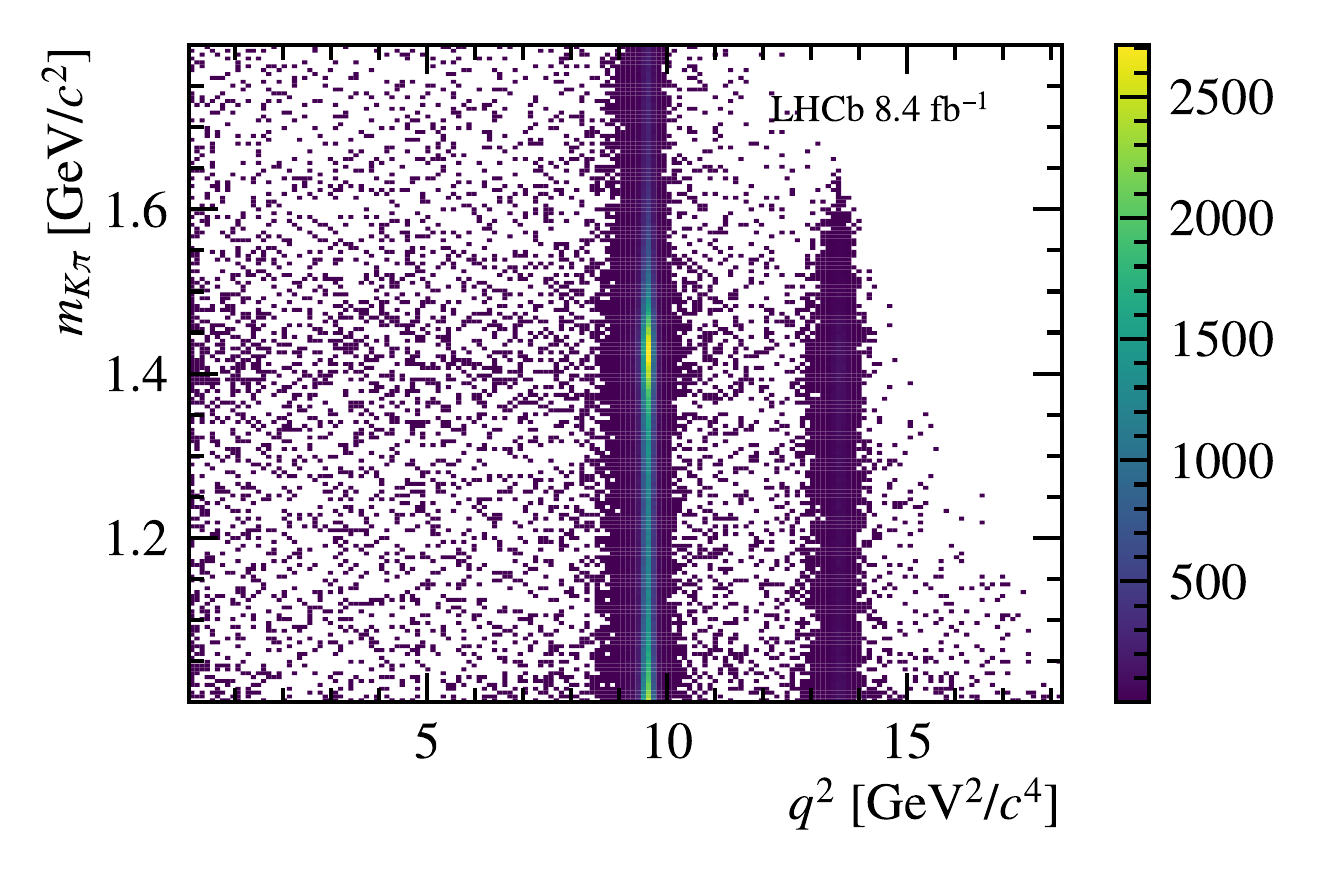}
    \includegraphics[width=0.47\textwidth]{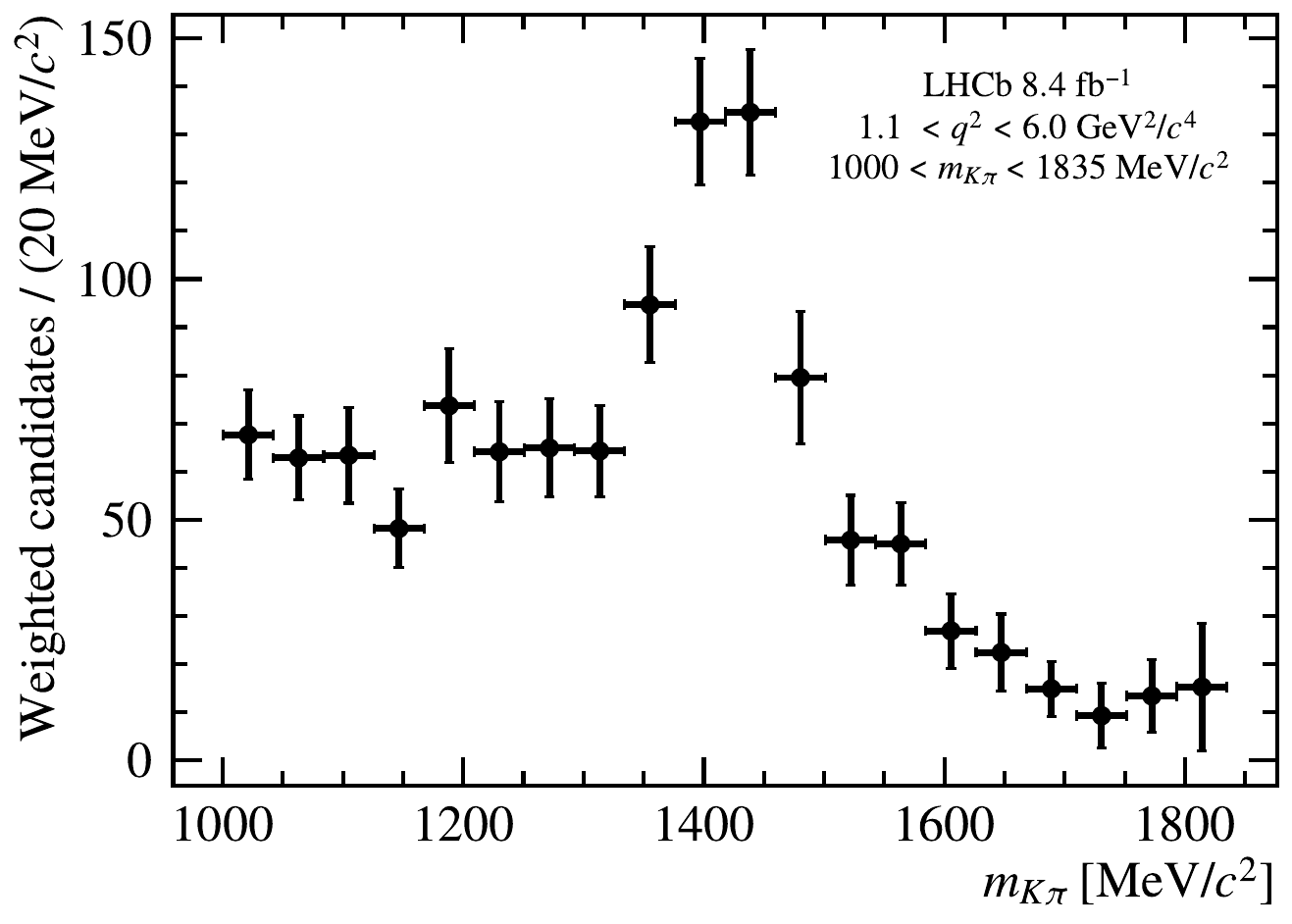}
    \caption{(Left) Two-dimensional $q^2$-$m_{K\pi}$ distribution.  The vertical bands correspond to the regions where the tree-level decays \decay{\Bd}{\jpsi K^+\pi^-} and  \decay{\Bd}{\psitwos K^+\pi^-}dominate. 
    (Right) Background-subtracted $m_{K\pi}$ distribution for signal candidates in the central-$q^2$ region, to which the $K^*_{0,2}(1430)$  S and D-wave resonances, as well as the P-wave $K^*_{1}(1410)$,  are expected to contribute. 
    }
    
    \label{fig:data-distribution}
\end{figure}

\label{{AppDists}}
Figure~\ref{fig:data-distribution} shows, on the left, the distribution of the selected data in the two-dimensional plane $q^2$-\mkpi, while, on the right, the background-subtracted distribution of \mkpi is shown for signal candidates in the central-$q^2$ region.
\clearpage

\clearpage 


\addcontentsline{toc}{section}{References}
\bibliographystyle{LHCb}
\bibliography{main,standard,LHCb-PAPER,LHCb-CONF,LHCb-DP,LHCb-TDR}

\newpage
\centerline
{\large\bf LHCb collaboration}
\begin
{flushleft}
\small
R.~Aaij$^{38}$\lhcborcid{0000-0003-0533-1952},
M.~Abdelfatah$^{69}$,
A.S.W.~Abdelmotteleb$^{57}$\lhcborcid{0000-0001-7905-0542},
C.~Abellan~Beteta$^{51}$\lhcborcid{0009-0009-0869-6798},
F.~Abudin\'en$^{59}$\lhcborcid{0000-0002-6737-3528},
T.~Ackernley$^{61}$\lhcborcid{0000-0002-5951-3498},
A.A.~Adefisoye$^{69}$\lhcborcid{0000-0003-2448-1550},
B.~Adeva$^{47}$\lhcborcid{0000-0001-9756-3712},
M.~Adinolfi$^{55}$\lhcborcid{0000-0002-1326-1264},
P.~Adlarson$^{87,42}$\lhcborcid{0000-0001-6280-3851}
% ...

{\footnotesize \it

% ...
$^{38}$Nikhef National Institute for Subatomic Physics, Amsterdam, Netherlands\\
% ...
$^{42}$National Center for Nuclear Research (NCBJ), Warsaw, Poland\\
% ...
$^{47}$Instituto Galego de F{\'\i}sica de Altas Enerx{\'\i}as (IGFAE), Universidade de Santiago de Compostela, Santiago de Compostela, Spain\\
% ...
$^{51}$Physik-Institut, Universit{\"a}t Z{\"u}rich, Z{\"u}rich, Switzerland\\
% ...
$^{55}$H.H. Wills Physics Laboratory, University of Bristol, Bristol, United Kingdom\\
% ...
$^{57}$Department of Physics, University of Warwick, Coventry, United Kingdom\\
% ...
$^{59}$School of Physics and Astronomy, University of Edinburgh, Edinburgh, United Kingdom\\
$^{60}$School of Physics and Astronomy, University of Glasgow, Glasgow, United Kingdom\\
$^{61}$Oliver Lodge Laboratory, University of Liverpool, Liverpool, United Kingdom\\
% ...
$^{69}$Syracuse University, Syracuse, NY, United States\\
% ...
$^{87}$Department of Physics and Astronomy, Uppsala University, Uppsala, Sweden, associated to $^{60}$\\
% ...
}
\end{flushleft}

\end{document}